\documentclass[12pt,a4paper]{article}
\usepackage{amssymb,amsmath}
\usepackage[dvips]{lscape,graphicx}

\newcommand{\ds}{\displaystyle}

\begin{document}

\vspace{2cm}

\title {\bf The Multiple Point Principle and Higgs Bosons}

\vspace{2cm}

\author{{\Large L.~V.~Laperashvili}\\[5mm]
Institute of Theoretical and Experimental Physics,\\[3mm]
B.Cheremushkinskaya str., 25, Moscow, Russia}

\date{}

\maketitle

\pagenumbering{arabic}

\begin{abstract}

The multiple point principle (MPP), according to which several
vacuum states with the same energy density exist, is put forward
as a fine-tuning mechanism predicting the ratio between the
fundamental and weak scales in the Standard Model (SM) and in its
two Higgs doublet extension (2HDM). Using renormalization group
equations for the SM, we obtain the effective potential in the
2-loop approximation and investigate the existence of its
postulated second minimum at the fundamental scale. In the SM an
exponentially huge ratio between the fundamental (Planck) and
electroweak scales results: $\frac{\Lambda_{fund}}{\Lambda_{ew}}
\sim e^{40}.$ But in the 2HDM the fundamental scale $\Lambda$ can
vary from 10 TeV up to the Planck scale. Using the MPP, we predict
the masses of the Higgs bosons.

\end{abstract}

\section {Cosmological Constant and the Multiple Point Principle}

In the present talk we suggest two scenarios:

I) The first scenario [1-4] considers only the pure Standard Model
(SM) with one Higgs boson. In this scenario we obtain an
exponentially huge ratio between the fundamental (Planck) and
electroweak scales: $\frac{\Lambda_{fund}}{\Lambda_{ew}} \sim
10^{17} \sim e^{40}$.

II) The second scenario [5] concerns to the general two Higgs
doublet extension of the SM, in which the fundamental scale
$\Lambda$ can vary from 10 TeV up to the Planck scale: $\Lambda
\sim 10^{19}$ GeV .

In such scenarios it is reasonable to assume the existence of a
simple postulate which helps us to explain the SM parameters:
couplings, masses and mixing angles. In our model such a postulate
is based on a phenomenologically required result in cosmology: the
cosmological constant is zero, or approximately zero, meaning that
the vacuum energy density is very small. A priori it is quite
possible for a quantum field theory to have several minima of the
effective potential as a function of its scalar fields.
Postulating zero cosmological constant, we are confronted with a
question: is the energy density, or cosmological constant, equal
to zero (or approximately zero) for all possible vacua or it is
zero only for that vacuum in which we live? This assumption would
not be more complicated if we postulate that all the vacua which
might exist in Nature, as minima of the effective potential,
should have approximately zero cosmological constant. This
postulate corresponds to the Multiple Point Principle (MPP)
developed in [6] (see also [7]): {\bf there are many vacua with
the same energy density or cosmological constant, and all
cosmological constants are zero, or approximately zero.}

Using the MPP, we predict the masses of the Higgs bosons.

 \section{The Higgs boson mass value in the SM}

 The success of the SM strongly supports the concept of
the spontaneous breaking of symmetry $SU(2)_L\times U(1)_Y \to
U(1)_{em},$ which is achieved by the Higgs mechanism, giving
masses of the gauge bosons $W^{\pm}$, $Z$, the Higgs boson and
fermions. This mechanism, in its minimal version, requires the
introduction of a single doublet of scalar complex Higgs fields
and leads to the existence of a neutral massive particle - the
Higgs boson.

Recently the experimental lower limit on the Higgs mass of 115.3
GeV was set by the unsuccessful search at LEPII. The energy
interval from 100 GeV to 200 GeV will be thoroughly examined at
the upgraded Tevatron, LHC and LC. These mashines have a good
chance to discover the Higgs boson in the near future.

With one Higgs doublet of $SU(2)_L$, we have the following
tree--level Higgs potential:
\begin{equation}        V^{(0)} = - m^2
\Phi^{+}\Phi + \frac{\lambda}{2}
        (\Phi^{+}\Phi )^2.
\end{equation} The vacuum expectation value of the Higgs field $\Phi$ is:
\begin{equation}               <\Phi> = \frac{1}{\sqrt 2}\left(
             \begin{array}{c}
             0\\
             v
             \end{array}              \right),
\end{equation} where
 $
 v = \sqrt{\frac{2 m^2}{\lambda}}\approx 246\,\,{\mbox{GeV}}.
$
The masses of gauge bosons $W$ and $Z$, fermions with flavor $f$
and the physical Higgs boson are given by the VEV parameter $v$:
\begin{equation}           M_W^2 = \frac{1}{4} g^2 v^2,
       \quad   M_Z^2 = \frac{1}{4} (g^2 + g'^2) v^2,
       \quad   m_f = \frac{1}{\sqrt 2} h_f v,
       \quad   M_H^2 = \lambda v^2,
\end{equation} where $h_f$ are the Yukawa couplings with flavor $f$.

In our paper [3] we have calculated the 2--loop effective
potential in the limit
      $\phi^2 >> v^2, \quad\quad \phi^2 >> m^2,$
using the SM renormalization group equations in the 2-loop
approximation [8].
Assuming the existence of the two minima of the effective
potential in the simple SM, we have taken the cosmological
constants for both vacua equal to zero, in accord with the
Multiple Point Principle.

\section{The Multiple Point Principle requirements}

The MPP requirements for the two degenerate minima in the SM are
given by the following equations:
\begin{equation}         V_{eff}(\phi_{min1}) = V_{eff}(\phi_{min2}) = 0,\quad
        V'_{eff}(\phi_{min1}) = V'_{eff}(\phi_{min2}) = 0,
\end{equation} \begin{equation}          V''_{eff}(\phi_{min1}) > 0, \quad\quad
          V''_{eff}(\phi_{min2}) > 0,
\end{equation} where
\begin{equation}          V'(\phi) = \frac{\partial V}{\partial \phi^2}, \quad
         \quad V''(\phi) = \frac{\partial^2
         V}{\partial(\phi^2)^2}.
\end{equation} As was shown in [1] the degeneracy conditions of MPP give the
following requirements for the existence of the second minimum in
the limit of high $\phi^2 >> m^2:$
\begin{equation}        \lambda(\phi_{min2}) = 0, \quad\quad {\mbox{and}}
       \quad\quad
 {\lambda'}(\phi_{min2}) = 0,
\end{equation} what means $
      \beta_{\lambda}(\phi_{min2}, \lambda=0) = 0.$

Using these requirements and the renormalization group flow,
C.D.Froggatt and H.B.Nielsen [1] computed quite precisely the top
quark (pole) and Higgs boson masses (see also the detailed
calculations in [3]):
\begin{equation} M_t = 173\pm 4 \,\,GeV \quad {\mbox{and}} \quad  M_H = 135\pm
9\,\, GeV. \end{equation}
\section{Two Higgs doublet extension of the SM}

There are no strong arguments for the existence of just a single
Higgs doublet, apart from simplicity. In the present talk we
consider the implementation of the MPP in the general 2HDM,
without any symmetries imposed beyond those of the SM gauge group.
The most general renormalizable $SU(2)\times U(1)$ gauge invariant
potential of the 2HDM is given by $$ V_{eff}(H_1, H_2)
=m_1^2(\Phi)H_1^+H_1+m_2^2(\Phi)H_2^+H_2- [m_3^2(\Phi)
H_1^+H_2+h.c.]$$ $$
+\ds\frac{\lambda_1(\Phi)}{2}(H_1^+H_1)^2+\frac{\lambda_2(\Phi)}{2}(H_2^+H
_2)^2+ \lambda_3(\Phi)(H_1^+H_1)(H_2^+H_2)^2$$ $$
+\lambda_4(\Phi)|H_1^{+}H_2|^2
+[\frac{\lambda_5(\Phi)}{2}(H_1^+H_2)^2+\lambda_6(\Phi)(H_1^+H_1)
(H_1^+H_2) $$ \begin{equation}
+\lambda_7(\Phi)(H_2^+H_2)(H_1^+H_2)+h.c.]\, , \end{equation}
where $$ H_i=\left( \begin{array}{c} \chi^+_i\\[2mm]
(H_i^0+iA_i^0)/\sqrt{2} \end{array} \right)\,. $$ The number of
parameters in the 2HDM compared with the SM grows from 2 to 10.

We suppose that mass parameters $m_i^2$ and Higgs self--couplings
$\lambda_i$ of the effective potential only depend on the overall
sum of the squared norms of the Higgs doublets, i.e.
\begin{equation}      \Phi^2=\Phi_1^2+\Phi_2^2\,,\qquad
\Phi_i^2=H_i^+H_i=(H_i^0)^2+(A_i^0)^2+|\chi_i^+|^2\,.
\end{equation}
The running of these couplings is described by the 2HDM
renormalization group equations [8], where the renormalization
scale is replaced by $\Phi$.

At the physical minimum of the 2HDM scalar potential the Higgs fields develop
vacuum expectation values (VEVs)
\begin{equation} <\Phi_1>=\ds\frac{v_1}{\sqrt{2}}\,,\qquad\qquad<\Phi_2>=\ds\frac{v_2}{\sqrt{2}},
\end{equation} breaking $SU(2)\times U(1)$ gauge symmetry and generating
masses of all bosons and fermions.

Here the overall Higgs norm: $
<\Phi>=\sqrt{v_1^2+v_2^2}=v=246\,\,{\mbox{GeV}} $ is fixed by the
electroweak scale. At the same time the ratio of the Higgs VEVs
remains arbitrary, and it is convenient to introduce $
\tan\beta=v_2/v_1.$

Assuming no CP, nor charge violation at the vacuum, the 2HDM
involves five physical states of the Higgs bosons:

1) The charged Higgs bosons $\chi^{\pm}$ have masses
$m_{H^{\pm}}$.

2) One CP--odd Higgs boson $A_2^0$ has a mass $m_A$.

3) The two CP--even scalars have masses $m_H$ and $m_h$.

The last one is a mass of the lightest Higgs particle.

\section{Implementation of the MPP in the 2HDM}

In this talk we present 2HDM supplemented by the MPP assumption.
We require that at some high energy scale
$
            M_Z << \Lambda \stackrel{<}{\sim } M_{Pl},$
which we shall refer as the MPP scale $\Lambda$,
the largest set of degenerate vacua is realized in 2HDM.
In compliance with the MPP, these vacua and the physical one
must have the same energy density.

We expected that the 2HDM effective potential, depending on the
norms $|H_1|$ and $|H_2| $, had to have two rings of minima in the
Mexican hat with the same vacuum energy density. The radius of the
little ring is at the electroweak scale $v$, while the radius of
the big one is $\Lambda $. In this case, imposing the MPP
conditions, we had to have that all the self-couplings should
vanish at the MPP scale: \begin{equation}
\lambda_1(\Lambda)=\lambda_2(\Lambda)=\lambda_3(\Lambda)=\lambda_4(\Lambda)=
  \lambda_5(\Lambda)=\lambda_6(\Lambda)=
  \lambda_7(\Lambda)=0\,.
\end{equation} But this case is in contradiction with the 2HDM
renormalization group equations for couplings. The position of
vacua depends on $\tan\beta $. In addition, the vacuum stability
conditions must be satisfied.

For example, there exists the MPP solution for the 2HDM vacua at
the scale $\Lambda$ which results in a set of degenerate vacua:
\begin{equation} <H_1>=\left( \begin{array}{c} 0\\[2mm] \Phi_1 \end{array} \right)\,,\qquad
<H_2>=\left( \begin{array}{c} \Phi_2\\[2mm] 0 \end{array}
\right)\,,\\[2mm]
\end{equation} where $ \Phi_1^2+\Phi_2^2=\Lambda^2,$ and
\begin{equation} \lambda_1(\Lambda)=\lambda_2(\Lambda)=\lambda_3(\Lambda)=0,\quad
\lambda_5(\Lambda)=\lambda_6(\Lambda)=\lambda_7(\Lambda)=0,
\end{equation} \begin{equation}
\beta_{\lambda_1}(\Lambda)=\beta_{\lambda_2}(\Lambda)=
\beta_{\lambda_3}(\Lambda)=0 \end{equation} with the vacuum
stability condition $ \lambda_4(\Lambda) > 0\,.$

This MPP solution allows to evaluate the Yukawa couplings of the
third generation at the electroweak scale in terms of only one
parameter $h_t(\Lambda)$. The values of $\tan \beta$ and
$h_t(\Lambda)$ can be fitted so that the correct values of the
top-quark and $\tau$-lepton masses are reproduced: $$
\begin{array}{c}m_t(M_t)=\ds\frac{h_t(M_t) v}{\sqrt{2}}\sin\beta\, , \qquad
m_{\tau}(M_t)=\ds\frac{h_{\tau}(M_t) v}{\sqrt{2}}\cos\beta\,.
\end{array} $$ Using experimentally given values: $$
  \alpha_3\approx 0.117, \quad m_t(M_t)\approx 165\,\, {\mbox{GeV}},
  \quad m_{\tau}(M_t)\approx 1.78\,\, {\mbox{GeV}},
$$ we obtain that $\tan \beta$ is confined around 50.

Finally the Higgs masses depend only on the two parameters: $
\Lambda   \quad {\mbox {and}} \quad m_A.$ We have made a detailed
numerical analysis of the MPP constraints on the Higgs spectrum in
the 2HDM for high energy scales ranging from $\Lambda=10$ TeV to
$\bf \Lambda=M_{Pl}$.

We present the following examples:

1) For $\Lambda = 10$ TeV and $m_A=400$ GeV, we have:\\
$ \tan\beta \approx 49.8,\quad
      m_h\approx 69\,\,{\mbox{GeV}},\quad
      m_{H^{\pm}}\approx 360\,\,{\mbox{GeV}},\quad
      m_H\approx 399\,\,{\mbox{GeV}}. $

2) For $\Lambda = 10^8$ GeV and $m_A=400$ GeV :
 $
       \tan\beta \approx 50,\quad
       m_h\approx 115\,\,{\mbox{GeV}},$
what corresponds to the LEPII experimental lower limit for the
light Higgs boson, and
$
       m_{H^{\pm}}\approx 387\,\,{\mbox{GeV}},
       \quad m_H\approx 399.5\,\,{\mbox{GeV}}. $

3) For $\Lambda = M_{Pl} = 1.22\cdot 10^{19}$ GeV and $ m_A=400$
GeV :\\ $
      \tan\beta \approx 50.3,\quad
      m_h\approx 137\pm 10\,\,\,{\mbox{GeV}},$
      what is quite close to the MPP prediction of the Higgs boson
      mass in the SM [1] (see Eq.(8)).
      Here we also have:
$
      m_{H^{\pm}}\approx 404\,\,{\mbox{GeV}},
      \quad m_H\approx 400\,\,{\mbox{GeV}}. $

We see that the masses of Higgs bosons grow with increasing
$\Lambda$.

In conclusion I want to emphasize that we tried to construct in
[5] a new simple MPP inspired non-supersymmetric two Higgs doublet
extension of the SM.

{\large \bf Acknowledgements:}

The speaker strongly thanks her co-authors: C.D.Froggatt,
R.B.Nevzorov, H.B.Nielsen and Marc Sher. She also thanks the
financial support by Russian Foundation for Basic Research,
project No.02-02-17379.


\begin{thebibliography}{99}

\bibitem{1}
C.D.Froggatt, H.B.Nielsen, Phys.Lett. B {\bf 368}, 96 (1996).
\bibitem{2}
C.D.Froggatt, H.B.Nielsen, L.V.Laperashvili, Hierarchy-Problem and
a Bound State of 6 $t$ and 6 $\bar t$. Invited talk by H.B.Nielsen
at the {\it Coral Gables Conference on Launching of Belle Epoque
in High-Energy Physics and Cosmology (CG2003), 17-21 Dec 2003,
Ft.Lauderdale, Florida, USA} (see Proceedings); ArXiv:
hep-ph/0406110.
\bibitem{3}
C.D.Froggatt, L.V.Laperashvili, H.B.Nielsen, The Fundamental-Weak
Scale Hierarchy in the Standard Model, to be published in Nucl.
Atom. Phys. (Yad. Fiz.); arXiv: hep-ph/0407102.
\bibitem{4}
C.D.Froggatt, L.V.Laperashvili, H.B.Nielsen, A New Bound State $6t
+ 6\bar t$ and the Fundamental-Weak Scale Hierarchy in the
Standard Model. A talk given by L.V.Laperashvili at the
International Seminar "QUARKS-2004", Pushkinskie Gory, Russia, May
2004; published in: Proceedings of the 13th International Seminar
on high energy physics", arXiv: hep-ph/0410243.
\bibitem{5}
C.D.Froggatt, L.V.Laperashvili, R.B.Nevzorov, H.B.Nielsen, M.Sher,
in preparation.
\bibitem{6}
D.L.Bennett, C.D.Froggatt, H.B.Nielsen, in {\it Proceedings of the
27th International Conference on High Energy Physics, Glasgow,
Scotland, 1994}, Ed. by P.Bussey and I.Knowles (IOP Publishing
Ltd, 1995), p.557; D.L.Bennett, H.B.Nielsen, Int.J.Mod.Phys. A
{\bf 9}, 5155 (1994).
\bibitem{7}
L.V.Laperashvili, Yad.Fiz. {\bf 57}, 501 (1994); Phys.Atom.Nucl.
{\bf 57}, 471 (1994).
\bibitem{8}
C.Ford, D.R.T.Jones, P.W.Stephenson, M.B.Einhorn, Nucl.Phys.{\bf
B395} (1993) 17.


\end{thebibliography}
\end{document}